\documentclass[aps,prl,twocolumn,showpacs,nobibnotes]{revtex4-1}
\usepackage{graphics,graphicx,amsfonts,amsmath,amsbsy,amssymb,color}
\usepackage[singlelinecheck=false]{subfig}
\usepackage{verbatim}  
\usepackage{psfrag}  
\usepackage{tabularx}
\usepackage{xmpmulti}

\def \bfq {{\bf q}}

\def \beq {\begin{eqnarray}}
\def \eeq {\end{eqnarray}}

\newcommand{\Hamil}{\hat{H}}

\newcommand{\reffig}[1]{{Fig.~\ref{#1}}}

\newcommand{\half}{\frac{1}{2}}

\begin{document}

\title{Correlation Energy Divergences in Metallic Systems}

\author{James~J.~Shepherd}
\email{js615@cam.ac.uk}
\affiliation{University of Cambridge, The University Chemical Laboratory, \\Lensfield Road, Cambridge, CB2 1EW, United Kingdom}
\author{Andreas~Gr\"{u}neis}
\email{andreas.grueneis@univie.ac.at}
\affiliation{University of Vienna, Faculty of Physics and Center for Computational Materials Science, Sensengasse 8/12, A-1090 Vienna, Austria}

\begin{abstract}
We numerically examine divergences of the total energy in metallic systems of approximate many-body theories using Hartree--Fock as a reference, including perturbative (M\o ller-Plesset, MP), coupled cluster (CC) and configuration interaction (CI) approaches. Controlling for finite size effects and basis set incompleteness error by comparison with energies from the random phase approximation (RPA), we suggest convincingly that non-perturbative coupled cluster theories are acceptable for modelling electronic interactions in metals whilst perturbative coupled cluster theories are not. Data are provided from the RPA with which it is possible to test other approximate finite-basis methods for divergences with only modest computational cost.
\end{abstract}
\date{\today}
\maketitle


\emph{Introduction.} -- Although it has long been known that finite order perturbation theory predicts divergent terms in the energy for metals, as analysed by Gell-Mann and Brueckner in 1957\cite{Brueckner1957},
it has taken fifty years for this result to be reproduced numerically for a real metallic system\cite{Grueneis2010}. This is in part because simulating the many-body wavefunction of solid state materials starting from Hartree--Fock theory is incredibly computationally expensive, often scaling polynomially with a variety of simulation parameters including the size of the simulation cell.
In spite of the difficulties in applying post-Hartree--Fock methods to solids, there is an increasing body of authors attempting to do so with a variety of wavefunction ansatzes on the so-called `hierarchies of quantum chemistry' 
\footnote{We have included a more pedagogical introduction to the literature in the Supplemental Materials}\cite{Paier2009,Dovesi1984,Gillan2008,Marsman2009,Schwerdtfeger2010,Schutz2007,Usvyat2011,Hirata2004,Grueneis2010,Scuseria2004,Hirata2004,Schwerdtfeger2010,Paulus2012,Manby2010,Grueneis2011,Schwerdtfeger2010,Paulus2012,Manby2010}. 
Their aim is to achieve highly accurate correlation energies for the field of materials science by leveraging the systematic improvability of these approximate solutions to the Schr\"{o}dinger equation\cite{Helgaker,Bartlett2007}.

An open question of growing importance surrounding this field is to directly address which methods are appropriate and which are not for the study of metallic systems\footnote{which a growing number of authors are attempting to address}. Approximations and divergences need to be understood so that needless effort is not expended investigating methods which will ultimately fail. Although it would in principle be possible to pursue this question with analytical theory, the plurality of diagrams and the lack of closed solutions makes this attempt intractable, especially for higher orders of coupled cluster theory.
So far most analytical results have been achieved by neglecting the exchange term in the Hartree--Fock reference which, in perturbative theories, physically corresponds to starting with a non-interacting reference. This greatly simplifies
the expression for the denominator in M\o ller-Plesset perturbation theory\cite{Plesset1934} and allows analytical divergences to be found\cite{Mattuck1967, Brueckner1957,Hirata2012}. Strictly speaking, however, these results consider only a lower bound to the true correlation energy
of the respective method using the Hartree--Fock reference.
On the other hand, one might think that a closing band gap is sufficient to guarantee divergences in a perturbative approach, however, it is well-known that some terms converge\cite{Onsager1966}. Furthermore, especially in real systems, numerical divergences can be hard to find and distinguish from convergences.
To this end we aim to provide here a simple, novel and robust methodology to test for the numerical convergence
of approximate methods in metals using the \emph{finite basis set} simulation-cell electron gas\cite{Shepherd2012a,Shepherd2012b,Shepherd2012c}. 

We focus in this study on testing a number of famous and commonly-used methods at the heart of quantum chemistry. These are coupled
cluster theory with full amplitude equations\footnote{In contrast to approximate amplitude equations\cite{Bishop1978,Bishop1982,Hirata2012}}\cite{Raghavachari1989,coester1958,coester1960,cizek1966,Bartlett2007}, M\o ller-Plesset perturbation theory\cite{Plesset1934} and truncated configuration
interaction (singles and doubles)\cite{Knowles2000,Meyer1976}.
These methods have been extremely successful in treating correlation in molecular systems and they are being increasingly applied to solid-state systems; extension to metallic systems is an ongoing aim in this community. 
We also draw
comparison with the random phase approximation\cite{Bohm1953} that has already received a great deal of attention from the solid-state physicists\cite{Harl2010,Schimka2010,Olsen2011}. These are only presented as examples, the methodology here is extendible to any wavefunction method which can be formulated in a finite canonical Hartree--Fock basis\cite{Hartree1928,Fock1930,Slater1930}.

\emph{Model.} -- The homogenous electron gas (uniform electron gas) 
consists of $N$ electrons in a box of length $L$ with a two-electron Ewald interaction $\hat{v}_{\alpha\beta}$\cite{Ewald1921,Fraser1996},
\begin{equation}
\Hamil=\sum_\alpha -\half \nabla_\alpha^2 + \sum_{\alpha\neq \beta} \half \hat{v}_{\alpha\beta} + \text{const.}
\label{sim_cell_H}
\end{equation}
In the thermodynamic limit, found as the particle number tends to infinity ($N\rightarrow\infty$) with the density held constant, it is possible to solve the above Hamiltonian in the Hartree--Fock Approximation with plane waves. This yields an analytic expression for the dispersion relation,
\begin{equation}
\epsilon_k = \half k^2 + \frac{k_F}{\pi} f\left(x \right)
\label{eq:ev}
\end{equation}
with
\begin{equation}
f\left(x \right)=\left( 1+\frac{1-x^2}{2x} \text{ln} \left| \frac{1+x}{1-x} \right| \right)
\end{equation}
and $x=k/k_F$, where this term is due to exchange\cite{Martin2004}. It is this term which produces a band structure with a zero in the density of states at the Fermi energy, and also complicates analytical derivations.

In setting out to find the behaviour of approximate theories to obtain the correlation energy (i.e. the total energy with Hartree--Fock energy as a starting-point), it is typical to start with a finite simulation-cell model of $N$ electrons, and carefully approach the thermodynamic limit by extrapolation\cite{Fraser1996,Drummond2008}. However, in quantum chemical techniques, we must also make do with a finite one-particle basis set. Very little has been studied about the relatively simple properties of the HEG represented in a finite basis, at least in part since the study of the electron gas has been pushed forward so much by continuum quantum Monte Carlo techniques that work at the complete basis set limit\cite{Ceperley1980,Foulkes2001,Needs2010}. The difficulty of investigating the properties of these approximate theories in the thermodynamic limit is hampered by this requirement of a finite basis set, in this case of $M$ plane waves spinorbitals defined by a kinetic energy cutoff $\frac{1}{2} k_c^2$. In principle, the complete basis set limit $k_c\rightarrow\infty$ and thermodynamic limit $N\rightarrow\infty$ must be found, which is prohibitively costly given the scaling of even approximate quantum chemical theories. 

The most obvious way to make progress towards these limits is to take the $k_c\rightarrow\infty$ limit, to solve the $N$-electron Hamiltonian at the complete basis set limit, and then the $N\rightarrow\infty$ limit can be found latterly. However, in this study we propose to take the $N\rightarrow\infty$ limit first for a finite $k_c$\footnote{Taking a \emph{finite-basis} approach to the homogeneous electron gas has only recently been introduced to the literature\cite{Shepherd2012a,Shepherd2012b,Shepherd2012c}}. We first outline how to show the well-known divergence in the MP2 energy using finite-$M$, finite-$N$ calculations and then generalise this approach to demonstrate limiting behaviours in other theories. 

In the thermodynamic limit, the 3D electron gas momentum space is a Fermi sphere containing all momenta less than the Fermi wavevector $k < k_F$. The virtual manifold of free-electron plane waves surround this sphere stretching out to infinite $k$ and infinitesimal spacing. In simulating this with finite systems, the finite electron number controls the spacing of the allowed plane-wave wavevectors and a finite basis set is produced by a cutoff, $\frac{1}{2}k^2\leq \frac{1}{2}k^2_c$, which controls the extent to which the plane waves stretch out into $k$-space around the $\Gamma$-point. Allowing $N\rightarrow\infty$ for a `finite-basis' electron gas is well-defined for a given $\gamma=k_c/k_F$ and amounts to representing the space inside $k\leq k_c$ with an ever finer grid. 

We can then simulate a series of $N$-electron gases with $M$ basis functions where $M = \gamma^3 N$ to within finite-size effects\footnote{this is technically achieved by allowing the energy cutoff defining the virtual-space to be proportionate to the the energy cutoff enclosing the occupied orbitals, which for small systems differs from the exact value of $k_F$, but is a more consistent treatment because both cutoffs are then minimum kinetic energy cutoffs}. For a given $\gamma$, as the $N\rightarrow\infty$ limit is taken, the band gap closes because the grid spacing in the region around the Fermi surface becomes smaller, and the zero-momentum ($\bfq={\bf 0}$) excitations that cause the divergences in for example MP2 theory are increasingly well-represented in a size-consistent fashion\footnote{A diagram and example table of $N$ and $M$ used here are provided in the Supplemental Materials to further illustrate these points.}.

\begin{figure}
\includegraphics[width=0.45\textwidth]{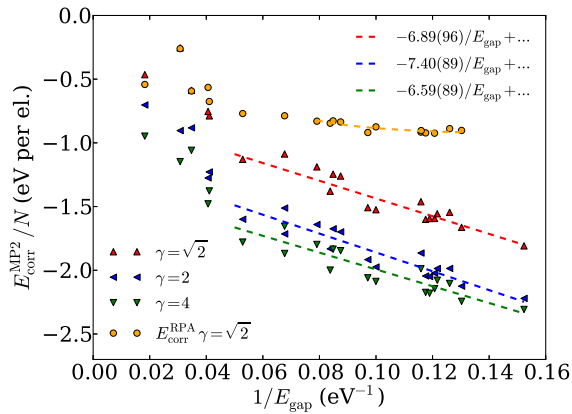}
\caption{(Colour online) MP2 energies for a variety of finite-basis simulation-cell electron gases with electron numbers $N=14-1030$ corresponding to closed-shell configurations of a simple cubic reciprocal-space lattice and density $r_s=1.0$~a.u. These all diverge as 1/$E_\text{gap}$ as the thermodynamic limit (TDL) is approached and the three different basis set sizes are parallel to within errors from the fit. For a single basis set size, the RPA correlation energy is also shown, which converges in the TDL.}
\label{fig:1}
\end{figure}

\emph{Results.} -- MP2 correlation energies are presented in \reffig{fig:1} for sets of finite-$N$, finite-$M$ electron gases constructed in this fashion. This conclusively demonstrates that this approach recovers the expected divergence and physical behaviour from this method which is insensitive to our choice of $\gamma$. In contrast, the differences between the different values of $\gamma$ converge to give a finite energy in the thermodynamic limit (TDL) represented by the lines for different $\gamma$ being parallel to one another within extrapolation error. Technically, this is just the basis set incompleteness error, which should be recovered as $1/\gamma^3$.
To further validate this as a physical approach accurately capturing the TDL, we compare this with the finite-basis electron gas energies from identically constructed RPA calculations, which show a convergent behaviour (with a finite-size error as $\sim N^{-1}$ \cite{Drummond2008}) as anticipated.
All RPA results in this work are calculated using a HF reference. 

\begin{figure}
\includegraphics[width=0.45\textwidth]{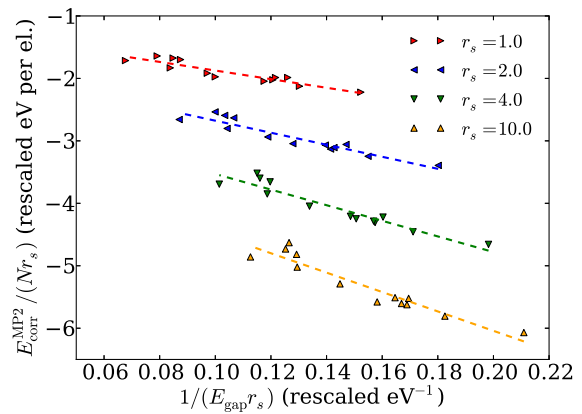}
\caption{(Colour online) MP2 energies for a single basis set size $\gamma=\sqrt{2}$ for $N=168-1030$ at a variety of densities showing that the divergent behaviour is insensitive to the density parameter. To compare on the same scale, the energies in both axes have been multiplied through by $r_s$, such that for example the true band gaps of the $r_s=10.0$~a.u. simulations span the energy range $\sim0.5-1.0$~eV.}
\label{fig:2}
\end{figure}

\begin{figure*}
  \subfloat[]{\includegraphics[width=0.45\textwidth]{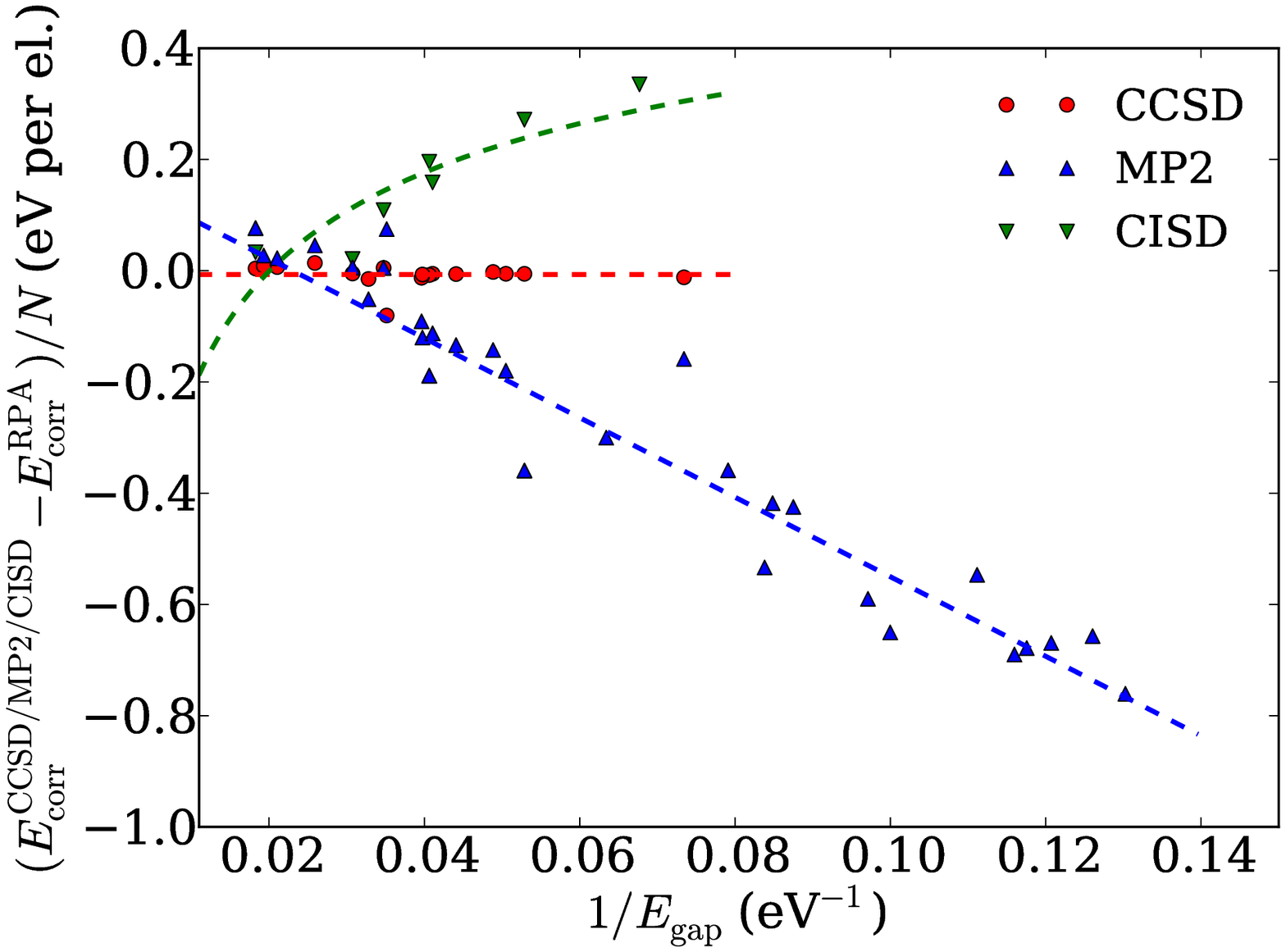}}
  \subfloat[]{\includegraphics[width=0.45\textwidth]{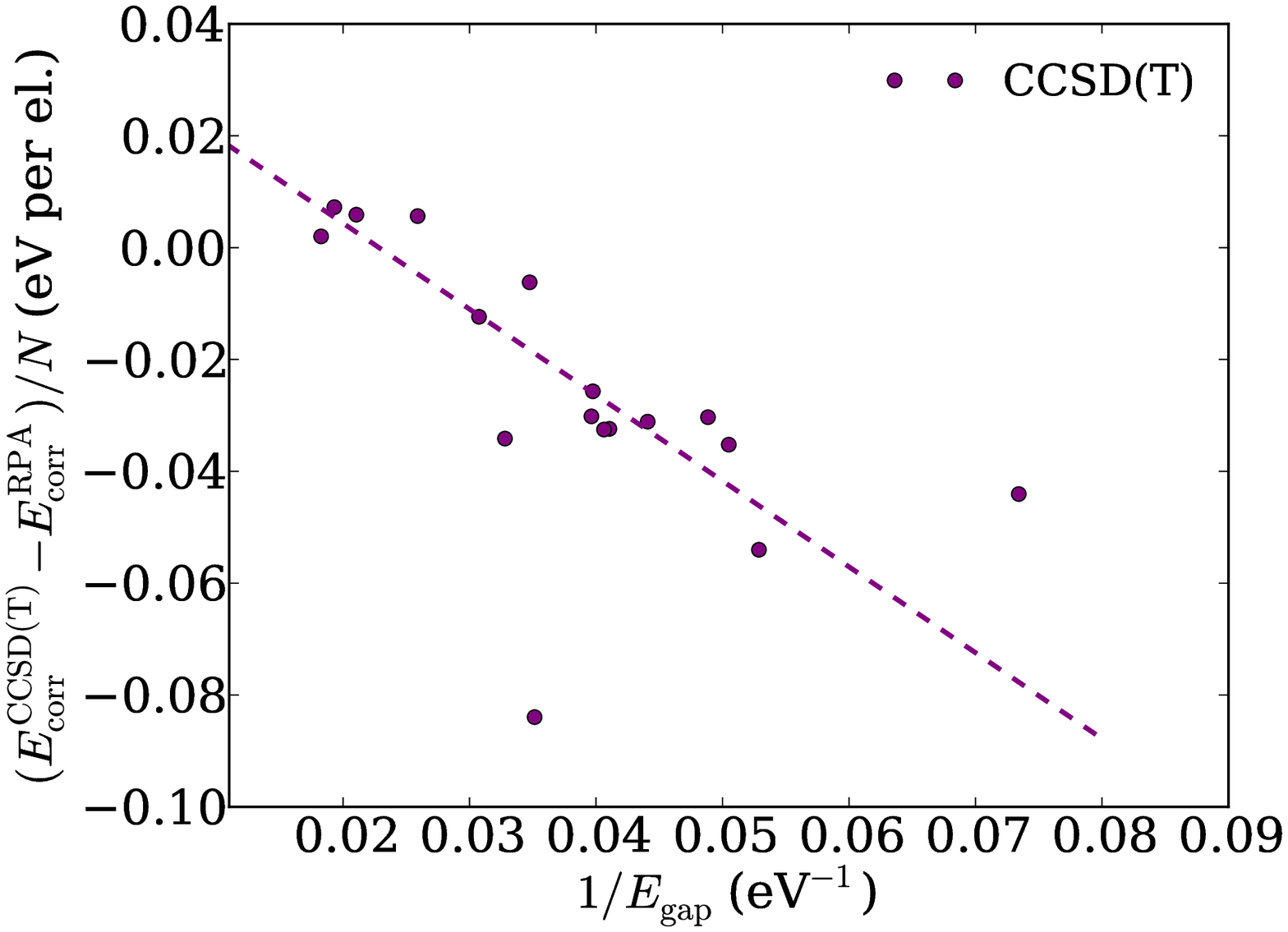}}
\caption{(Colour online) Energy differences between the RPA correlation energy ($r_s=1.0$~a.u., $\gamma=\sqrt{2}$) show that MP2 and CCSD(T) are divergent as the band gap closes (on approach to the thermodynamic limit). CCSD is convergent to a constant energy offset with respect to RPA which is only serendipitously close to zero for this $r_s$ and $\gamma$. CISD gradually yields zero correlation energy per electron, due to a lack of size-extensivity.}
\label{fig:3}
\end{figure*}  

We further note that the band gaps of the simulation-cell electron gas in these simulations, at $r_s=1.0$ (relatively high density), are still fairly large ($54.8-6.6$ eV) due to remaining finite-size errors even though we are simulating electron numbers between $N=14-1030$. To examine the insensitivity of the divergent behaviour with respect to a wider range of band gaps, we have compared simulations of several densities (by changing the $r_s$ parameter) over the same range of electron numbers in \reffig{fig:2}. As the density is lowered and $r_s$ is raised, the range of typical metallic densities of metals is traversed\cite{Mermin1996}, and although different contributions of kinetic and exchange energy are present in the Hartree--Fock energy, the divergence is remarkably unchanged. Furthermore, we now see that the divergence persists at a very large range of simulation-cell band gaps. This underlines a main conclusion of this study, that in spite of simulating what appear to be insulating finite systems, the divergence in the MP2 energy is still visible and the energy is without even qualitative physical relevance.

Having now numerically demonstrated the well-accepted behaviour for the MP2 energy, we turn our attention towards other approximate methods for which the behaviour on approach to the TDL is unverified --- coupled cluster singles and doubles theory (CCSD) and the addition of perturbative triples (CCSD(T)). There has been surprisingly little literature concerning this hierarchy of methods as applied to solids, in spite of the wealth of applications they have received in the molecular quantum chemistry community.
Even though there has been some discussion of CCSD with \emph{approximate} amplitude equations, these more closely resemble the RPA equations\cite{Freeman1977,Bishop1978,Bishop1982,Hirata2012}.
As such, to the best of our knowledge, the question of whether CCSD and CCSD(T) diverge in the TDL for metallic systems has not yet been conclusively addressed and requires further investigation. We note in passing that CCSD and coupled cluster doubles theory (CCD) are equivalent for the homogeneous electron gas due to the complete absence of symmetry-allowed single-excitations in its many-body expansion.

Due to the relatively expensive scaling of such methods, simulations of a $N=1030$ electron gas with current fully-periodic codes\cite{VASP} are prohibitively expensive. However, we have found that further reduction in finite-size effects can be achieved by taking the difference between the CCSD or CCSD(T) energy with the RPA energy and in this difference the limiting behaviour is more clear. We have also taken advantage of other simulation-cell lattices (face-centered cubic and body-centered cubic) to provide more closed-shell configurations. Taking energy differences in this way allows us to clearly demonstrate, in \reffig{fig:3}, that the CCSD energy is strongly convergent, whereas the CCSD(T) diverges at the same speed as the MP2 energy\footnote{The outliers at $\sim$0.035 and $\sim$0.075 eV$^{-1}$ are due to these (bcc) systems having spuriously low and high band gaps respectively due to the lattice shape.}. 
%

\begin{figure*}
\includegraphics[width=0.75\textwidth]{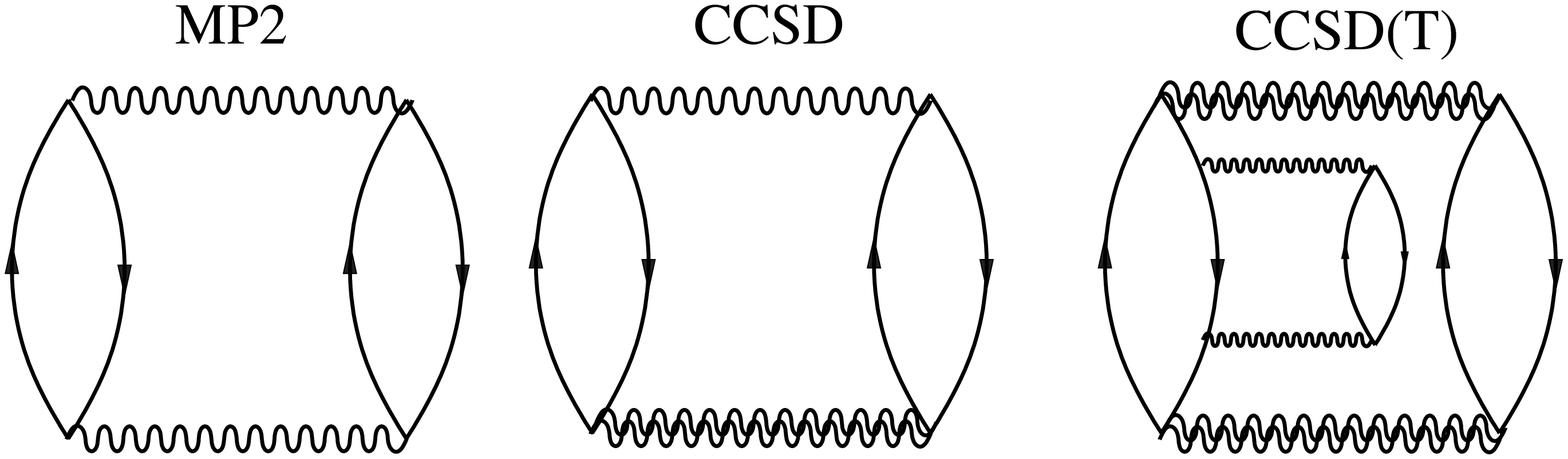}

\begin{minipage}[b]{0.05\linewidth} \centering \ \end{minipage}\begin{minipage}[b]{0.25\linewidth} \centering \begin{equation} E^{(2)}_d=\sum_{ijab} \frac{2 \langle ij | ab \rangle \langle ab | ij \rangle }{ \Delta\epsilon_{ijab}}\notag \end{equation} \begin{equation} E^{(2)}_d \approx \int \frac{1}{{\bf q}^4 \epsilon_g({\bf q})} d {\bf q}\notag \end{equation} \end{minipage}\begin{minipage}[b]{0.25\linewidth} \centering \begin{equation} E^\text{CCSD}_d=\sum_{ijab} {2 \langle ij | ab \rangle t_{ij}^{ab} }\notag \end{equation} \begin{equation} E^\text{CCSD}_d \approx \int \frac{ V_\text{eff}({\bf q})}{{\bf q}^2 } d {\bf q}\notag \end{equation} \end{minipage}\begin{minipage}[b]{0.3\linewidth} \centering \begin{equation} E^\text{(T)}_d = \sum_{ijkabcd} \frac{ t_{ik}^{ac} \langle aj | db \rangle \langle db | aj \rangle {t_{ik}^{ac}}^* }{\Delta\epsilon_{ijkabc}}\notag \end{equation} \begin{equation} E^\text{(T)}_d \approx \int \int { V^2_\text{eff}({\bf q})} \frac{1}{{\bf p}^4 \left( \epsilon_g ({\bf p}) +\epsilon_g({\bf q})+\epsilon_g({\bf q}+{\bf g}) \right) }d {\bf p} d {\bf q} \notag \end{equation} \end{minipage}\begin{minipage}[b]{0.05\linewidth} \centering \ \end{minipage}

\caption{Goldstone diagrams present in MP2, CCSD and CCSD(T). Below the diagrams the respective algebraic expressions are given and approximated for the uniform electron gas assuming a Fermi point. $i,j,k$ and $a,b,c,d$ refer to occupied and unoccupied orbitals, respectively. $\Delta\epsilon_{ijab}$ and $\epsilon_g$ refer to the differences between the occupied and unoccupied one-electron energies. $V_{\rm eff}$ denotes a non-singular screened Coulomb interaction. The exact form of $V_{\rm eff}$ depends on the form of the coupled cluster amplitude equations. In the RPA $V_{\rm eff}$ is given by $\epsilon^{-1}({\bf q})/{\bf q}^2$, where $\epsilon({\bf q})$ is the dielectric matrix.}

\label{fig:4}
\end{figure*}

For a better understanding of the above results we may consider the dominant Goldstone diagrams included in the respective methods.
\reffig{fig:4} shows the diagrams responsible for the divergences in MP2 and CCSD(T) as well as an diagram
contributing to the CCSD energy. The horizontal (double) wavy lines correspond to the (screened) Coulomb interactions.
The vertical lines denote occupied and unoccupied orbitals. 
Algebraic expressions given beneath these diagrams correspond to the HEG assuming a point-like Fermi sphere\footnote{in common with other authors, e.g. \onlinecite{Mattuck1967}}. 
Evidently, the unscreened Coulomb interactions are singular for zero momentum-transfer vectors.
In combination with the closing Hartree--Fock (HF) gap, this is the reason for the divergence in the MP2 energy.
In contrast, for CCSD, our numerical results suggest that the effectively screened Coulomb interaction is non-singular
and leads to a finite CCSD correlation energy in metals.
The effective non-singular Coulomb interaction is obtained by a summation of infinitely many diagrams, which in the RPA correspond to the so-called ring diagrams and in CCSD is carried out by solving the amplitude equations\footnote{Although there are variants to CCSD in which approximations are made to the amplitude equations. In this work we employ no approximations to the CCSD amplitude equations.}.
CCSD(T), however, includes a fourth-order diagram that corresponds algebraically to an expression that involves an integration
over two unscreened Coulomb interactions and a closing HF gap. This expression closely resembles the MP2 expression.
As such, the CCSD(T) correlation energy diverges at the same rate as the MP2, albeit with a smaller
prefactor as is corroborated by our numerical findings.

For completeness, we also demonstrate the effect of this difference when the test method is not size-extensive. As an example of this, we used configuration interaction singles and doubles (CISD, and equivalent to CID for the HEG), obtained by using a truncation of a CI quantum Monte Carlo calculation\cite{Booth2009,Cleland2010,Thom2010,Clark2012}, and this shows a tendency to retrieve an ever-lower fraction of the correlation energy as the electron number is raised (\reffig{fig:3}). A simple rationalisation of this is as follows.
When two $N$-particle systems are individually treated with CID, they achieve a better treatment of correlation than the combined 2$N$-particle system since in the individual systems, the consideration of quadruple excitations in the larger system is made absent \cite{Knowles2000}. Due to this, correlation energy is not recovered consistently as $N$ grows but, per particle, falls as some inverse polynomial.
In particular, correlation energy is lost as $N^{-\half}$ in the case of non-interacting electron pairs\cite{Szabo1996,Helgaker}.

\emph{Concluding remarks.} -- In summary, we have shown that a judicious choice of finite-size and finite-electron number homogeneous electron gas models
can be used to demonstrate the limiting behaviour of the correlation energy in approximate many-body theories for
metallic systems with modest computational cost. By comparing to RPA correlation energies we control for basis set incompleteness and finite-size errors.
As a first application of the outlined methodology, we have verified the divergence of MP2 energies in metals.
Moreover, we have shown that CCSD(T) also exhibits a divergent behaviour in the HEG, by virtue of the perturbative
(T) correction to the CCSD correlation energy.
In contrast, CCSD, due to an ``effectively screened'' Coulomb interaction, predicts converging correlation energies in metals.
Our findings strongly suggest that the cutting-edge accuracy of high-level \emph{perturbative} quantum chemistry methods such as CCSD(T)
will only be transferable from molecular quantum chemistry to the study of metallic solids, if novel theories
are introduced that lift the divergent behaviour.
This work provides a simple but stringent test for such novel many-body approximations that are too complex
for analytical theory, and we have provided data such that other authors may conduct the same analysis\footnote{See Supplemental Material for raw data including RPA correlation energies.}.

\emph{Acknowledgements.} -- The authors thank Ali Alavi, Georg Kresse and Alex J. W. Thom for discussions. One of us (AG) gratefully acknowledges an APART-fellowship of the Austrian Academy of Sciences, and the other (JJS) EPSRC for funding.


\bibliography{divergencesbib}

\clearpage
\appendix

\begin{table*}
\centering
\newcolumntype{R}{>{\centering\arraybackslash}X}
\begin{tabularx}{\textwidth}{|R|R|R|R|R|R|R|R|R|}
\hline
\multicolumn{3}{| >{\setlength\hsize{3\hsize}\centering\arraybackslash}R|}{Method (Acronym)} & Formal scaling & Exact through & Applied to solids & Size-extensive energies  \\
\hline
\hline
\multicolumn{3}{| >{\setlength\hsize{3\hsize}\centering\arraybackslash}R|}{Hartree--Fock Theory (HF) \cite{Hartree1928,Fock1930,Slater1930} }  & $N^3$ & 1st order & \onlinecite{Paier2009,Dovesi1984,Gillan2008,Marsman2009} & Yes \\
\hline
\multicolumn{3}{| >{\setlength\hsize{3\hsize}\centering\arraybackslash}R|}{Second-order Moller-Plesset Theory (MP2) \cite{Plesset1934} }  & $N^5$ & 2nd order & \onlinecite{Schwerdtfeger2010,Schutz2007,Usvyat2011,Hirata2004,Grueneis2010,Scuseria2004} & Yes  \\
\hline
\multicolumn{3}{| >{\setlength\hsize{3\hsize}\centering\arraybackslash}R|}{Random phase approximation (RPA) \cite{Bohm1953}} & $N^4$ & 1st order & \onlinecite{Harl2010,Schimka2010,Olsen2011} & Yes  \\
\hline
\multicolumn{3}{| >{\setlength\hsize{3\hsize}\centering\arraybackslash}R|}{Coupled cluster doubles with approximated amplitude equations} & - & approx. 3rd order & \onlinecite{Freeman1977,Bishop1978,Bishop1982} & Yes \\
\hline
\multicolumn{3}{| >{\setlength\hsize{3\hsize}\centering\arraybackslash}R|}{Coupled cluster singles and doubles theory (CCSD) \cite{coester1958,coester1960,cizek1966}} & $N^6$ & 3rd order & \onlinecite{Hirata2004,Schwerdtfeger2010,Paulus2012,Manby2010,Grueneis2011} & Yes  \\
\hline
\multicolumn{3}{| >{\setlength\hsize{3\hsize}\centering\arraybackslash}R|}{Coupled cluster singles and doubles theory with perturbative triples (CCSD(T))\cite{Raghavachari1989}} & $N^7$ & 4th order & \onlinecite{Schwerdtfeger2010,Paulus2012,Manby2010}  & Yes  \\
\hline
\multicolumn{3}{| >{\setlength\hsize{3\hsize}\centering\arraybackslash}R|}{Configuration interaction truncated at single and double excitations (CISD)\cite{Meyer1976}} & $N^6$ & - & - & No  \\
\hline
\end{tabularx}
\caption{(Supplemental Material.) Pedagogical summary of the acronyms, references and scaling of the canonical formulation of the employed many-body theories. We note that the scaling of these methods might be naturally lower for the HEG due to momentum symmetry constraints on the allowed excitations, but this is \emph{not} considered here. References to solid-state applications are intended as examples rather than an exhaustive list.}
\label{tab:refs}
\end{table*}

\begin{figure*}
\includegraphics[width=0.5\textwidth]{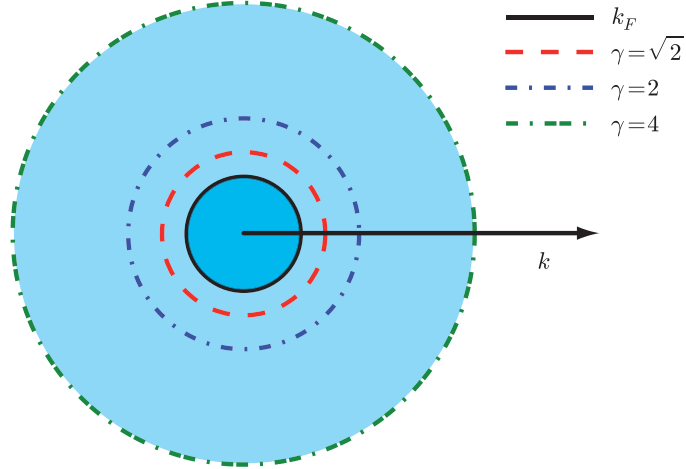}
\caption{(Colour online, supplemental material.) Scale diagram illustrating various finite-basis TDL electron gases at a variety of $\gamma$. The dark shaded circle indicates the location of the fermi sphere, and hence wavevectors of occupied orbitals, and the light section indicates wavevectors of virtual orbitals.}
\label{diag}
\end{figure*}

\begin{table*}
\centering
\newcolumntype{R}{>{\centering\arraybackslash}X}
\begin{tabularx}{\textwidth}{R R R R R}
\hline
Electron number $N$ & Plane wave spinorbitals $M$ & HF band gap $E_\text{gap}$ (eV)& Correlation energy $E_\text{corr.}^\text{RPA} $(eV / electron) \\
\hline
\hline
  14   & 38    &     54.752 & -0.541 \\   
  38   & 66    &     32.521 & -0.260 \\ 
  54   & 162   &     28.776 & -0.593 \\
  66   & 186   &     24.357 & -0.675 \\
  114  & 294   &     18.911 & -0.770 \\
  162  & 358   &     24.614 & -0.565 \\
  186  & 514   &     14.773 & -0.787 \\
  246  & 682   &     12.642 & -0.830 \\
  294  & 778   &     11.932 & -0.846 \\
  342  & 922   &     11.787 & -0.827 \\
  358  & 970   &     11.428 & -0.836 \\
  406  & 1174  &     10.297 & -0.919 \\
  514  & 1502  &     8.6289 & -0.915 \\
  610  & 1694  &     8.5071 & -0.922 \\
   682 &  1850 &     8.4075 & -0.904 \\
   730 &  2042 &     8.2846 & -0.924 \\
   778 &  2090 &     7.6773 & -0.903 \\
   874 &  2378 &     7.9345 & -0.888 \\
   922 &  2474 &     10.002 & -0.874 \\
\hline
\end{tabularx}
\caption{(Supplemental Material.) RPA correlation energies and band gaps for the systems electron gas models studied here ($r_s=1.0$~a.u., $\gamma=\sqrt{2}$). Other lattice types are available on request by email.}
\label{tab:1}
\end{table*}

\end{document}